\documentclass[aps,prl,floatfix,10pt,twocolumn,superscriptaddress,letterpaper,showpacs]{revtex4-1}
\usepackage{graphicx,bm,color,amsmath,amssymb,amsfonts,mathrsfs,txfonts}

\newcommand{\al}{\alpha}
\newcommand{\be}{\beta}
\newcommand{\g}{\gamma}

\newcommand{\z}{\zeta}

\newcommand{\thi}{\theta}

\newcommand{\mi}{\mu}

\newcommand{\p}{\pi}

\newcommand{\s}{\sigma}

\newcommand{\f}{\phi}

\newcommand{\pd}{\partial}

\newcommand{\round}[1]{\left( #1 \right)}
\renewcommand{\square}[1]{\left[ #1 \right]}

\newcommand{\cvec}[2]{\round{\begin{array}{c} #1 \\ #2 \end{array}}}

\newcommand{\mat}[4]{\left(\begin{array}{cc}#1&#2\\#3&#4\end{array}\right)}

\newcommand{\ang}[1]{\left\langle #1 \right\rangle}

\newcommand{\beq}{\begin{equation}}
\newcommand{\eeq}{\end{equation}}
\newcommand{\Beq}{\begin{eqnarray}}
\newcommand{\Eeq}{\end{eqnarray}}
\newcommand{\bml}{\begin{multline}}

\newcommand{\bsp}{\begin{split}}
\newcommand{\esp}{\end{split}}

\newcommand{\nn}{\nonumber}

\newcommand{\tf}{\tilde\phi}

\newcommand{\ve}{\varepsilon}

\begin{document}
\title{Noise due to neutral modes in the $\nu=2/3$ fractional quantum Hall state}
\author{So Takei}
\affiliation{Department of Physics and Astronomy, University of California, Los Angeles, California 90095, USA }
\author{Bernd Rosenow}
\affiliation{Physics Department, Harvard University, Cambridge, Massachusetts 02138, USA}
\affiliation{Institut f{\"u}r Theoretische Physik, Universit{\"a}t Leipzig, D-04103, Leipzig, Germany}
\author{Ady Stern}
\affiliation{Department of Condensed Matter Physics, Weizmann Institute of Science, Rehovot 76100, Israel}
\date{\today}

\begin{abstract}
We theoretically study charge noise generated by excited neutral modes, which impinge on the quantum point contact of a  quantum Hall bar with filling fraction $\nu=2/3$. The noise is computed for thermally excited neutral modes as well as for biased neutral modes with dipole-fermion excitations.
Within the dipole-fermion picture, we show that the noise arising from two colliding  modes 
can be suppressed due to Pauli-blocking and be non-universal due to random edge disorder, but becomes universal upon 
 disorder-averaging. The ratio of noise due to two colliding neutral modes and noise due to only one such mode is 
 smaller for dipole-fermions  than  for thermal excitations, thus providing evidence for the  different quantum statistics of the two types of excitations.

\end{abstract}

\maketitle

The behavior of two-dimensional (2D) electron gases in the fractional quantum Hall (FQH) 
regime has garnered much attention ever since its discovery~\cite{FQHEexp}. While FQH liquids are 
incompressible with finite energy gaps to all bulk excitations, they
support one or more 1D gapless conducting channels along their boundaries~\cite{fqhedge}. 
The edges of some FQH states, such as the spin-polarized $\nu=2/3$~\cite{kf23} and 
the $\nu=5/2$ Pfaffian and anti-Pfaffian non-Abelian states~\cite{pfapf1,pfapf2}, are predicted to possess neutral 
modes which can carry heat but no charge. The Majorana degree of freedom in the (anti-)Pfaffian neutral mode
is essential for the non-Abelian statistics of these states. For the case of $\nu=2/3$, our focus in this paper, the neutral mode flows opposite to the charge mode. While the existence of neutral modes was first predicted 
almost 20 years ago~\cite{kf23,kfprb}, experimental evidence for their existence
was obtained only recently 
using shot noise 
measurements~\cite{nmexp1,nmexp2,nmexp3} and quantum dot thermometers~\cite{nmexp4,nmexp5}.

In an abelian FQH state with two edge channels, the neutral mode is the dipole excitation of the composite edge | made of charges with opposite signs on the two edge channels. 
  Theoretically, for  the specific example of the random 2/3-edge the neutral mode is made of spinful chiral fermions ~\cite{kf23,kfprb}. Its excitations may be expressed as spin flips of these fermions. In the presence of disorder, the number of these excitations is not conserved, since the scattering of an electron from the $\nu=1$ mode to the counter-propagating $ \nu=1/3$ mode creates such a spin flip excitation in the neutral mode. 
While neutral mode heat transport can be partially understood in terms of thermaly excitated bosonic collective modes, we focus here on possible manifestations of the fermionic nature of the neutral mode. We refer to the fermions as ``dipole fermions". 

In this letter, we address the question how neutral mode 
dipole-fermion excitations  and their interaction with a QPC can be described theoretically, and we identify signatures of their quantum statistics  in a  setup where two biased neutral modes collide at a QPC. By drawing on an analogy with the effect of a current bias on edge correlation functions of the charge mode, we associate an oscillatory behavior of neutral mode edge correlation functions 
with a biased neutral mode. To quantify our findings, we introduce the ratio $\gamma$ between the current noise at a QPC due to two colliding neutral modes and the current  noise due to only one impinging neutral mode. We find that for biased neutral modes,  $\gamma$  is not only   smaller than two, but it is also  much smaller than for the case of colliding thermally excited neutral modes. This result 
is a fingerprint of quantum statistics. It originates from a partial Pauli-blockade of occupied states,  and also allows to experimentally distinguish dipole-fermion  excitations from thermaly excitated neutral modes. Our heuristic description of biased neutral modes is backed up by a full-fledged Keldysh calculation, 
and a comparison of our theoretical results with the recent experiment \cite{nmexp3} suggests that the collision of thermally excited neutral modes was observed there.

\begin{figure}[t]
\centering
\includegraphics[scale=0.3]{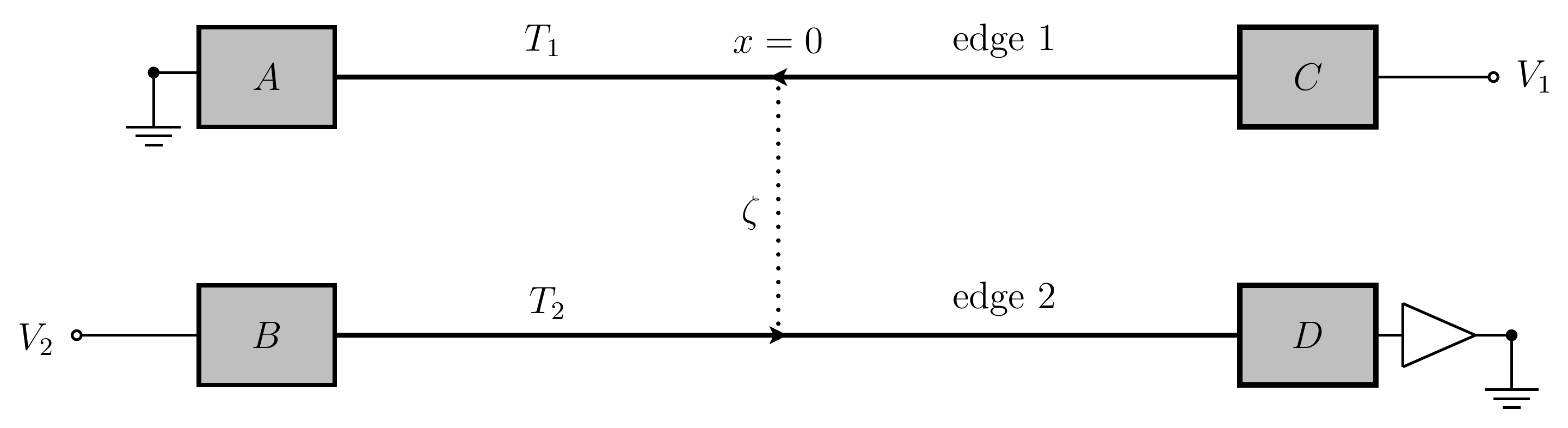}
\caption{\label{fig:edges} A quantum Hall bar with a single charge mode on each edge. The upper and lower edges are labeled 1 and 2, respectively. 
The tunneling amplitude between the edges is given by $\z$.}
\end{figure}

{\em General discussion of shot noise versus thermal noise }: 
Before delving into detailed calculations, 
we first present a simple interpretation of noise generation due to thermal excitations and a current bias, respectively. 
For now, it suffices to consider a QH bar in the $\nu =1$ quantum Hall  state  as shown in Fig.~\ref{fig:edges}, 
and focus on the dc current noise arising at the QPC due to stochastic backscattering of quasi-particles there. 
In order to make connection with the discussion of fractional edges to follow, 
the modes on the two edges are described 
by two oppositely chiral boson fields with Lagrangian densities  
$\mathcal{L}_j=\pd_x\phi_j((-)^{j-1}\pd_t-u\pd_x)\phi_j/2$, where $j=1,2$ labels the edges and $u$ denotes
the edge plasmon velocity. 
Backscattering at the QPC can be modeled by the tunneling term 
$\mathcal{L}_{\rm tun}=  {T} + T^\dagger$, with 
%
\begin{equation}
{T} \ = \ e^{i e V t}  \z e^{i\varphi_1(x=0,t)-i\varphi_2(x=0,t)} \ \ ,
\end{equation}
which describes the tunneling of  charge $ e$ electrons. The exponential factor containing the bias $V$ between upper and lower edge describes the time evolution of electron operators on the upper edge versus those on the lower edge. The  current backscattered at the QPC is  given by 
$I_B = -i e \hbar^{-1} ({T} - {T}^\dagger)$, and  current noise is then obtained from the current-current correlation function 
 $S_B(t)=\ang{I_B(t)I_B(0)}$ via $S_B=\int_{-\infty}^\infty dt S_B(t)$.
 %
%
Computing the current correlation function to leading order in the backscattering, the expectation value of the 
two current operator factorizes into a product of correlation functions of the quasi-particle operator evaluated on the same edge, 
\beq
\label{sbsimple}
S^0_B(t)=e^2|\z|^2 \cos(e  V t) F_1(T_1,t)   F_1(T_2,t)/2
\eeq
where $F_{1} (T_j,t) = \langle e^{i \varphi_j(x=0,t)} e^{-i \varphi_j(x=0,0)} \rangle$ is the finite temperature ``greater" correlation function of the electron operator.

When the edge modes are  at zero temperature and zero bias the dc current noise vanishes. We now show that this happens due to a cancellation of short- and long-time contributions. 
Under these conditions,  $F_{1}(T_j=0,t)=\tau_c/(\tau_c+it)$, where $\tau_c$ is a short-time cutoff. 
The dc noise has a singular contribution from short-time correlations,
$S_{B,\, {\rm short}}^{0}=  \int_{-\ve}^\ve dt S_B^0(t)$, and another from long-time edge correlations
 $S_{B, \, {\rm long}}^{0}= (\int_{-\infty}^{-\ve}+\int_{\ve}^\infty) dt S_B^0(t)$. Here, $\ve$ has to be taken on the order of $\tau_c$, 
 but $\ve > \tau_c$ by a numerical factor. 
 In the ground state of the system, the noise vanishes because of a precise cancellation between these two contributions, $S_{B,\, {\rm short}}^{0} + S_{B,\, {\rm long}}^{0} =0$. This can be seen from contour integration in the lower half plane, using a closed contour consisting of the real axis and an ``infinite"  semicircle around the origin. Since the contour does not contain a singularity, the integral vanishes. Due to the long-time decay $\propto 1/t^2$ of the integrand, the integral over the infinite semicircle vanishes, and we obtain the 
 desired cancellation of $S_{B,\, {\rm short}}^{0}$ and   $S_{B,\, {\rm long}}^{0}$. 
 
If the long-time behavior of the current correlation function is modified,  finite noise can result. In the following, we consider two mechanisms for such a modification, one due to finite temperature and another due to a finite bias voltage. At a finite temperature $T > 0$ there is an exponential suppression of long-time correlations according to 
%
\beq
\label{sbsimplet}
F_1(T,t)= {\p  k_BT \tau_c/\hbar \over \sin[(\p  k_BT/\hbar)(\tau_c+it)]} \ \ .
\eeq
%
For $t\sim 0$, the function essentially remains unchanged from the zero temperature form, but the long-time (i.e. $t>\hbar/k_BT$)
correlations are now suppressed exponentially.  This suppresses the long-time contribution to noise, 
$S_{B,\, {\rm short}} > -S_{B,\, {\rm long}}$, and non-zero dc noise results. 

A finite bias causes an oscillatory behavior of the tunneling operator Eq.~(1), and as a consequence gives rise to  a temporally-oscillating factor to (\ref{sbsimple}), 
%
\begin{equation}
S_B(t)=S^0_B(t)\cos(\mu t/\hbar) \ \ .
\label{bias}
\end{equation}
%
The singular contribution $S_{B,\, {\rm short}}$ coming from $t\sim 0$ is again unaffected by the oscillatory 
factor, but the oscillations suppress the non-local time correlations and lead as before to non-zero noise. 
Later on, we will describe the edge correlation function of  a biased neutral mode by such an oscillatory factor as well, and 
relate the frequency of the oscillations to the magnitude of the neutral current. We find that for the case of two excited neutral modes impinging on a QPC from opposite sides, the noise contributions of thermally excited neutral modes add up, reflecting the bosonic character of thermal excitations, while there is a suppression of noise for dipole fermions that impinge from two biased edges at zero temperature, again reflecting the quantum statistics of excitations.

We now extend the above interpretation for noise generation to the $\nu=2/3$ FQH edge. Let us consider impinging
an excited neutral mode only on the upper edge. In the presence of a neutral heat current, one assumes that the charge and 
neutral modes on the edge are fully equilibrated at the QPC. The heated neutral mode raises the temperature of its
partner charge mode until the two equilibrate at a common temperature which is above the lower edge base temperature.
In analogy with the above discussion, this exponentially suppresses temporal correlations on the upper edge and generates charge 
noise at the QPC. This thermal mechanism for noise was the basis behind the heat transport picture~\cite{nmheat2} previously 
developed to explain the experiment~\cite{nmexp1} consisting of a single excited neutral mode. Further below, 
the noise  will be rigorously evaluated in the presence of thermally excited neutral modes on {\em both} edges.

We may also consider a model for a biased neutral mode, which formally introduces 
a temporally-oscillating factor in the correlation function for the neutral quasi-particles, just as in Eq. (\ref{bias}). 
This model corresponds to neutral dipole-fermion excitations, which are the main focus of the present work.
Since only the neutral mode is biased, we introduce the oscillatory factor to the 
correlation function for the neutral mode only and leave the charge sector unperturbed. 
One may na{\"i}vely expect that 
biased neutral modes alone should not generate {\em charge} noise at the QPC. However, since
the most relevant tunneling operators for the $\nu=2/3$ edge involve tunneling of quasi-particles which are
superpositions of both the charge and neutral modes, an excited neutral mode indeed generates charge noise
at the QPC within the dipole-fermion model as well. 

It is important to note that neutral dipole-fermions need not be conserved during tunneling, unlike
charged Laughlin quasi-particles which are subject to charge conservation. This means that
the tunneling Hamiltonian for dipole-fermions allows terms that correspond to the creation or destruction
of two quasi-particles on both edges. Ignoring the charge modes for the moment, the Hamiltonian modeling
the tunneling of the neutral dipole-fermions can then be written as 
\beq
\label{htunn}
H_{\rm tun}=\z_1 e^{i\varphi_{\s1}(x=0)-i\varphi_{\s2}(x=0)}+\z_2 e^{i\varphi_{\s 1}(x=0)+i\varphi_{\s 2}(x=0)}+h.c.\,,
\eeq 
where $\varphi_{\s j}$ denotes the phonon field corresponding to the neutral mode on edge $j$.
Introducing the oscillatory factors as before via $e^{i\varphi_{\s j}(x=0)}\rightarrow e^{i\varphi_{\s j}(x=0)}e^{-i\mu_{\s j} t/\hbar}$
should generally lead to two types of oscillatory factors in the backscattering noise
\beq
\label{sbsimpledp}
S_B(\tau)\propto S^0_B(\tau)\square{\cos(\mu_{\s}^-\tau/\hbar)+\cos(\mu_{\s}^+\tau/\hbar)},
\eeq
where $\mu_\s^\pm=\mu_{\s 1}\pm\mu_{\s2}$, we have assumed $\z_1=\z_2$ for simplicity,
and $\mu_{\s j}$ denotes the bias voltage on edge $j$. The first term in the square brackets, with dependence on 
$\mu_\sigma^-$, describes the behavior expected for fermions: due to the Pauli principle, there is no noise if the bias voltages on the two edges are equal. The second term with dependence on  $\mu_\sigma^+$ arises due to the fact that the number of dipole-fermions is not conserved in scattering processes. The presence of  both these terms  leads to the generation of finite noise for equally biased neural modes, however with a reduced power  as compared to the case of thermally excited neutral modes. The reduction of 
noise power is due to the first term reflecting the quantum statistics of dipole fermions, thus making the reduced noise power of dipole fermions as compared to bosonic thermal excitations  a signature of the quantum statistics of neutral dipole-fermions. 
In the following, we substantiate this result  with a more rigorous microscopic calculation. 







{\em Dipole picture for noise in the random $\nu=2/3$ state}: In the standard theory for the
random $\nu=2/3$ FQH edge, interactions and disorder drive the edge to a disorder-dominated fixed-point
where the edge reconstructs into a weakly-coupled charge mode propagating downstream and a counter-propagating neutral mode~\cite{kf23}. 
At the fixed-point, the neutral mode possesses an exact SU(2)-symmetry,
and the propagation of the neutral mode can be interpreted as a flow of dipole (i.e. spinor) fermions. 
The edge disorder randomly \textit{rotates} the quantization  axes of the dipoles as they propagate
spatially along the edge. 


Let us consider a $\nu=2/3$ QH bar with a single QPC located at
$x=0$. The real-time action for the random $\nu=2/3$ edges
is given by $S=\sum_{j=1,2}[S_{\rho j}+S_{\s j}+S_{{\rm int},j}+S_{{\rm rand},j}]$, where 
$S_{\rho j}=(4\p\nu)^{-1}\int dt dx\ \pd_x\f_{\rho j}[(-)^{j-1}\pd_t-u_\rho\pd_x]\f_{\rho j}$,
$S_{\s j}=(8\p)^{-1}\int dt dx\ \pd_x\f_{\s j}[(-)^{j}\pd_t-u_\s\pd_x]\f_{\s j}$,
$S_{{\rm int},j}=(u_{{\rm int},j}/4\p)\int dt dx\ \pd_x\f_{\rho j}\pd_x\f_{\s j}$, and
%
\begin{equation}
S_{{\rm rand},j}=\int dt dx\ [\xi(x)e^{i\f_{\s j}}+c.c.]  \ \ , 
\end{equation}
%
which describes the spatially random rotations of the dipole quantization axes associated with the neutral modes. 
Here, $\f_{\rho j}$ and $\f_{\s j}$ are the charge and neutral phonon modes on edge $j$, and the coupling between these modes
is given by $u_{{\rm int},j}$. 
The random edge potential $\xi(x)$ is uncorrelated
on lengths scales longer than the mean-free path $\ell_0$. 
The neutral sector $S_{\s j}+S_{{\rm rand},j}$ 
is characterized by a level-one SU(2) current algebra spanned by $e^{\pm i\f_\s}$ and $\pd_x\f_\s$, which transform
as $S_\pm$ and $S_z$ in an SU(2) algebra. Therefore, for $u_{{\rm int},j}=0$, one can map the neutral sector to
an SU(2)-invariant fermionic action written in terms of a two-component spinor $\Psi$, and the random term can then be solved using a
space-dependent random SU(2) rotation $\tilde\Psi(x)=U(x)\Psi(x)$, where $U(x)$ again is uncorrelated on length scales beyond
$\ell_0$.  
Although a finite $u_{{\rm int},j}$ breaks the SU(2)-symmetry, the presence of the random potential
renders $u_{{\rm int},j}$ irrelevant in the RG sense, and the charge-neutral decoupled random fixed point is stable.

The quasi-particle tunneling at the QPC is described by the three most relevant tunneling operators~\cite{kf23},
\beq
\label{qpops}
e^{i(\f_{\rho}-\f_{\s})/2}, \ e^{i(\f_{\rho}+\f_{\s})/2}, \ e^{i\f_\rho},
\eeq
where the first two correspond to quasi-particles with charge $e/3$ and the last with charge $2e/3$.
As discussed above, we describe the case of  biased dipole fermions  by multiplying the first two tunneling operators Eq.~(\ref{qpops})
containing the neutral phonon fields
$\mp \phi_{\s j}$ with oscillatory terms $\exp( \pm i\mu_jt/\hbar)$, thus modeling the generation of noise due to a neutral current bias.

Disorder randomly rotates the dipole quantization  axes of the neutral modes over spatial length scales 
set by $\ell_0$. The noise generated by the neutral modes will depend on the
orientations of the polarization axes at the QPC.
Since $e^{\pm i\f_\s}$ represent spin-one excitations $S_\pm$ in an SU(2) algebra, the
$e^{\pm i\f_\s/2}$ create spin-half objects and form a basis for the 2D
representation for SU(2). At the tunneling site, $e^{\pm i\f_\s/2}$  then transforms as 
\beq
\cvec{e^{i\f_{\s j}/2}}{e^{-i\f_{\s j}/2}}=\mat{e^{i\varphi_j}\al_j}{e^{i\varphi_j}\be_j}{-e^{-i\varphi_j}\be_j}{e^{-i\varphi_j}\al_j}
\cvec{e^{i\tf_{\s j}/2}}{e^{-i\tf_{\s j}/2}},
\eeq
where $\al_j=\cos(\thi_j/2)$ and $\be_j=\sin(\thi_j/2)$, and $\thi_j$ and $\varphi_j$ are the polar and azimuthal angles for edge $j$ at the QPC. 
After implementing this rotation on the tunneling operators (\ref{qpops}), 
the backscattering correction to the dc charge noise can be obtained using standard methods 
\begin{multline}
\label{SBr}
S_B(\{\thi_j\},\{\varphi_j\},\mu_1,\mu_2)=\frac{(e^*)^2}{8\hbar^2}\int dt\left[(|\z_{11}|^2+|\z_{21}|^2)
\right.\\\times\cos[(\mi_{1}-\mu_{2})t/\hbar]
\left.+(|\z_{12}|^2+|\z_{22}|^2)\cos[(\mi_{1}+\mu_{2})t/\hbar]+4\z_3^2\right]
\\\times F_{2g}(T_0,t)[1-(2itk_BT_0/\hbar)].
\end{multline}
Here, $\z_{11}=\z_1e^{-i\bar\varphi}\al_1\al_2+\z_2e^{i\bar\varphi}\be_1\be_2$, 
$\z_{12}=\z_1e^{-i\bar\varphi}\al_1\be_2-\z_2e^{i\bar\varphi}\be_1\al_2$,
$\z_{21}=\z_1e^{-i\bar\varphi}\be_1\be_2+\z_2e^{i\bar\varphi}\al_1\al_2$, 
$\z_{22}=\z_1e^{-i\bar\varphi}\be_1\al_2-\z_2e^{i\bar\varphi}\al_1\be_2$, $\bar\varphi=\varphi_1-\varphi_2$
and $F_{2g}(T,t)=((\p  k_BT\tau_c/\hbar)/\sin(\p  k_BT/\hbar)(\tau_c+it))^{2g}$. 
Here, $g=2/3$, 
and $\z_i$ is the tunneling matrix element associated with the $i$-th quasi-particle in (\ref{qpops}), and
the base temperature is denoted by $T_0$. 
The excess noise is then defined as usual by $S_{\rm ex}(\{\thi_j\},\{\varphi_j\},\mi_1,\mi_2)
=S_B(\{\thi_j\},\{\varphi_j\},\mi_1,\mi_2)-S_B(\{\thi_j\},\{\varphi_j\},0,0)$. Note that (\ref{SBr})
verifies the schematic result (\ref{sbsimpledp}).

If both dipole quantization axes are along the $z$-axis and the neutral modes on both
edges are excited equally, i.e. $\thi_j=0$ and $\mu_1=\mu_2=\mu$, we see from (\ref{SBr})
that the excess noise vanishes. This is a manifestation of Pauli-blocking which arises from the fermionic nature of the underlying excitations.
If the neutral mode on the two edges are asymmetrically excited (say $\mu_1>\mu_2\ge0$) finite excess noise
results, and the maximal excess noise is obtained when only one edge is excited.

\begin{figure}[t]
\centering
\includegraphics*[scale=0.4]{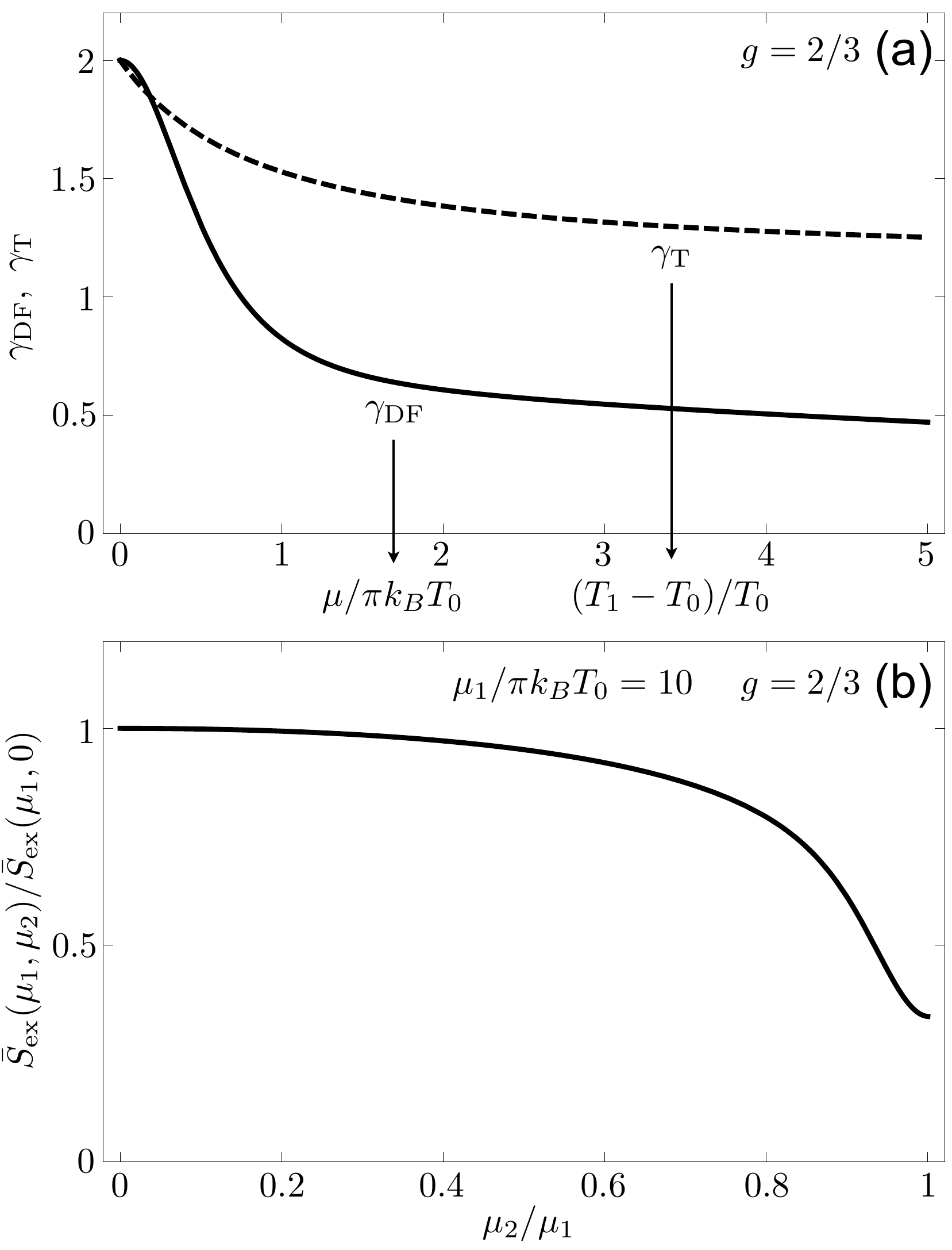}
\caption{\label{fig:ratio} (a) Excess noise ratio for one and two excited neutral modes within the dipole picture $\g_{\rm DF}$ (solid line) and the thermal picture $\g_{\rm T}$ (dashed line). (b) Excess noise ratio for one and two excited neutral modes within the dipole picture while fixing the edge 1 bias in the asymptotically large regime, $\mu_1/\p k_BT_0=10$, and varying the edge 2 bias, $\mu_2$.}
\end{figure}

If the spatial region over which tunneling occurs at the QPC is much less than $\ell_0$, 
the resulting excess noise is non-universal since it depends sensitively on the
axis orientations, $\thi_j$ and $\varphi_j$, at the tunneling site.
However, in the opposite limit, one is justified to average the excess noise
over these angles, and we arrive at a universal value for the excess noise ratio.
Averaging over both rotations, $\bar S_B(\mi_1,\mi_2)=(1/4\p)^2\prod_{j=1}^2\int_0^{2\p}d\varphi_j\int_0^{\p}\sin\thi_jd\thi_j S_B$,
we obtain
\begin{align}
\label{SDC_CD}
\bar S_B(\mi_1,\mi_2)=&\frac{(e^*)^2}{8\hbar^2}\int dt\left[(\z_{1}^2+\z_{2}^2)
\cos(\mi_{1}t/\hbar)\cos(\mu_{2}t/\hbar)\right.\nn\\
&\qquad\left.+4\z_3^2\right]F_{2g}(T_0,t)[1-(2itk_BT_0/\hbar)].
\end{align}
Transforming the product of the two $\cos$-factors in the first line of the equation into a sum, the dependence on oscillation rates 
$\mu_{1/2}/\hbar$ is exactly the same as that in Eq.~(\ref{sbsimpledp}), thus giving support to the intuitive picture discussed above. 
We now define the disorder-averaged excess noise, 
$\bar S_{\rm ex}(\mi_1,\mi_2)=\bar S_B(\mi_1,\mi_2)-\bar S_B(0,0)$. 
We compare two cases: (i) single excited neutral mode, i.e. $\mi_1=\mu$ and $\mi_2=0$; and (ii) two excited neutral modes, 
i.e. $\mi_1=\mi_2=\mi$. The ratio, $\g_{\rm DF}=\bar S_{\rm ex}(\mu,\mu)/\bar S_{\rm ex}(\mu,0)$,
is shown as the solid line in Fig. \ref{fig:ratio}(a). 
We have taken $\hbar/k_B\tau_c=0.5$K and $T_0=10$mK, which are consistent with Ref.~\onlinecite{nmexp3},
and we have also taken $\z_1=\z_2=\z_3=\z$. A clear signature of noise suppression due to quantum statistics can be seen in Fig.~\ref{fig:ratio}(b). Here, the bias for the dipole fermions on edge 1 is fixed at a large value, i.e. $\mu_1/\p k_BT_0=10\gg1$, and the bias on edge 2 is varied from $0$ to $\mu_1$. We interpret the suppression of noise as $\mu_2\rightarrow\mu_1$ as due to the reduction in phase space for scattering at the QPC from Pauli blocking. The suppression does not reach strictly zero due to disorder-averaging.

{\em Thermal picture for noise for the random $\nu=2/3$ state}: In this case, the backscattering correction to the dc noise reads
\begin{multline}
\label{SDC_HT}
S_B(T_1,T_2)=\frac{(e^*)^2}{8\hbar^2}\int dt\left[\z_{1}^2+\z_{2}^2+4\z_3^2\right]
\\\times F_{g}(T_1,t)F_{g}(T_2,t)[1-(2itk_BT_0/\hbar)],
\end{multline}
where $T_1, T_2\ge T_0$ are the temperatures of the upper and lower edges at the QPC, which are 
heated above the base temperature $T_0$ via transport of heat by the neutral modes~\cite{nmheat2}. 
We define the normalized noise as above, and the excess noise is defined as $S_{\rm ex}(T_1,T_2)=
S_B(T_1,T_2)-S_B(T_0,T_0)$. We consider again two cases: (i) a single thermally-excited
neutral mode, i.e. $T_1>T_0$ and $T_2=T_0$; and (ii) two thermally-excited neutral modes, i.e. 
$T_1=T_2>T_0$. 
The ratio, $\g_{\rm T}=S_{\rm ex}(T_1,T_1)/S_{\rm ex}(T_1,T_0)$,
 shown as the dashed line in Fig. \ref{fig:ratio}(a), is considerably higher than for the dipole-fermion case.

{\em Comparison to experiment}: 
For $\mu/\p k_BT_0\gg1$ [see solid line in Fig.~\ref{fig:ratio}(a)], 
which may be the regime of relevance for Ref.~\onlinecite{nmexp3}, 
we see that the theoretical prediction for $\g_{\rm DF}$ falls well below the experimental value $\g_{\rm exp}\approx1.6$. 
We have checked that this result does not depend on specific values for $\z_i$.
For thermally excited neutral modes [see dashed line in Fig.~\ref{fig:ratio}(a)], we see that the experimental value is reproduced 
when $(T_1-T_0)/T_0\approx 0.6$. Ref.~\onlinecite{nmexp3} provides lower and upper 
estimates, $T_1=55$mK and $T_1=200$-300mK, for the effective 
temperature at the QPC, putting either of these estimates well above the
estimated base temperature of $T_0=10$mK. As $(T_1-T_0)/T_0\to\infty$ the asymptotic value approaches $\g_{\rm T}\approx1.3$, 
lower than the asymptotic value $\gamma_{\rm exp}$ found in \cite{nmexp3}.
However, treating the scaling dimension $g$ of the edge correlation function as a  parameter of the theory as in \cite{RoHa02,PaMa04},  we have reevaluated the noise for $g=1$. 
In the dipole-fermion picture we find that the theoretical ratio
remains well below the experimental value for $\mu/\p k_B T_0\gg1$, while in the thermal picture we now find  
$\g_{\rm T}\approx 1.5$ in the limit of strong thermal bias 
$(T_1-T_0)/T_0 \approx 10$, in better agreement with the experiment \cite{nmexp3}, and indicating that 
thermally excited neutral modes are observed there. 

In conclusion, we studied  the excess charge noise generated by impinging one or two excited neutral modes on
a single QPC. We focused on the $\nu=2/3$ FQH state, and considered thermally excited neutral modes as well as 
biased neutral modes with  dipole-fermion excitations. We have found that in the case of two neutral modes impinging on a QPC the noise power reflects the quantum statistics of bosonic thermal excitations vs. dipole-fermions, thus allowing to detect a fingerprint of 
the quantum statistics of neutral edge excitations. 

{\em Acknowledgments $-$} BR acknowledges the DFG for financial support. AS thanks Microsoft Station Q, the US-Israel BSF and the Minerva Foundation for financial support.

\bibliography{Noise_Dip_Pic}

\end{document}